\documentclass[aps,pra,twocolumn,preprintnumbers,showpacs,superscriptaddress,amsmath,amssymb]{revtex4}
\usepackage{dcolumn}
\usepackage{bm}
\usepackage{amssymb}
\usepackage[german, english]{babel}
\usepackage{graphicx}
\usepackage{color}
\usepackage{textcomp}
\usepackage[letterpaper,total={7in,9.5in},top=0.75in,left=0.75in]{geometry}

\usepackage{array}

\begin{document}
\title{Measuring the $^{229}$Th nuclear isomer transition with $^{233}$U-doped crystals}

\author{Simon Stellmer}
\affiliation{Vienna Center for Quantum Science and Technology (VCQ) and Atominstitut, TU Wien, 1020 Vienna, Austria}
\author{Matthias Schreitl}
\affiliation{Vienna Center for Quantum Science and Technology (VCQ) and Atominstitut, TU Wien, 1020 Vienna, Austria}
\author{Georgy Kazakov}
\affiliation{Vienna Center for Quantum Science and Technology (VCQ) and Atominstitut, TU Wien, 1020 Vienna, Austria}
\author{Johannes Sterba}
\affiliation{Vienna Center for Quantum Science and Technology (VCQ) and Atominstitut, TU Wien, 1020 Vienna, Austria}
\author{Thorsten Schumm}
\affiliation{Vienna Center for Quantum Science and Technology (VCQ) and Atominstitut, TU Wien, 1020 Vienna, Austria}

\date{\today}

% 06.30.Ft Time and frequency
% 78.20.-e Optical properties of bulk materials and thin films

\begin{abstract}
We propose a simple approach to measure the energy of the few-eV isomeric state in $^{229}$Th. To this end, $^{233}$U nuclei are doped into VUV-transparent crystals, where they undergo alpha decay into $^{229}$Th, and, with a probability of 2\,\%, populate the isomeric state. These $^{229\rm{m}}$Th nuclei may decay into the nuclear ground state under emission of the sought-after VUV gamma, whose wavelength can be determined with a spectrometer. 

Based on measurements of the optical transmission of $^{238}$U:CaF$_2$ crystals in the VUV range, we expect a signal at least 2 orders of magnitude larger compared to current schemes using surface-implantation of recoil nuclei. The signal background is dominated by Cherenkov radiation induced by beta decays of the thorium decay chain. We estimate that, even if the isomer undergoes radiative de-excitation with a probability of only 0.1\%, the VUV gamma can be detected within a reasonable measurement time.

\end{abstract}

\maketitle

	% Introduction

    % 229Th is exciting, the nuclear clock is one possible application
The nucleus of $^{229}$Th is believed to possess an extremely low-lying excited nuclear state with an energy of only a few eV \cite{Kroger1976fot,Reich1990eso,Burke1990aef,Helmer1994aes,Beck2007eso}. The notion of applying well-established techniques of optical laser spectroscopy to this nuclear system has encouraged a great number of proposals for possible applications \cite{Dykhne1998aac,Peik2003nls,Flambaum2006eeo,Rellergert2010cte,Tkalya2011pfa,Campbell2012sin,Das2013qie}, among them the development of a nuclear optical clock \cite{Peik2003nls,Campbell2012sin,Kazakov2012poa}.

    % gamma spectroscopy, next step would be optical spectroscopy with a synchrotron and a crystal
Beginning in the 1970s, a series of gamma spectroscopy measurements with ever increasing performance used a differencing scheme to determine the energy of the $J^P[Nn_z\Lambda] = 3/2^+[631]$ isomeric state in an indirect way. The latest measurement places the energy at 7.8(5)\,eV above the $5/2^+[633]$ ground state \cite{Beck2007eso,Beck2009ivf}, corresponding to a transition wavelength of 160(10)\,nm. Predictions of the lifetime of the unperturbed isomeric state range between a few minutes and a few hours \cite{Helmer1994aes,Dykhne1998meo,Ruchowska2006nso}, where a value of around 1000\,s for a bare nucleus is the most commonly used estimate.

A logical next step in refining the transition wavelength, as well as demonstrating the optical addressibility of this nuclear two-level system, would be optical spectroscopy employing synchrotron radiation on $^{229}$Th nuclei doped into, or adsorbed onto \cite{Jeet2015roa,Yamaguchi2015esf}, vacuum ultraviolet (VUV)-transparent crystals. Such crystals allow one to place $10^{16}$ nuclei into the excitation beam, compared to a maximum number of $10^{6}$ particles in an ion trap \cite{Campbell2011wco,Herrera2014elo}.

    % this approach suffers from a number of drawbacks
The approach of direct optical excitation faces a number of severe challenges: (1) The quantity of $^{229}$Th (half-life $t_{1/2} = 7932(55)\,$a \cite{Kikunaga2011dot}) available for research is very limited, and growing crystals with high Th doping concentrations is a challenge \cite{Jeet2015roa}. (2) Both the lifetime and the energy of the isomeric state are known only with large uncertainties \cite{Beck2007eso,Sakharov2010ote,Jeet2015roa}, requiring an extensive two-dimensional search. (3) The choice of sufficiently tunable light sources in the VUV range is essentially limited to synchrotrons, where beam time is precious, yet the spectral power is some 11 orders of magnitude smaller compared to diode lasers in the visible range. (4) Photoluminescence in response to the excitation light can persist for long times, potentially masking the nuclear signal \cite{Stellmer2015rap}. (5) De-excitation of the isomeric state may proceed on parasitic pathways such as internal conversion (IC) and coupling to electronic states and phonons of the crystal, potentially suppressing the optical de-excitation altogether \cite{Tkalya2000dot,Karpeshin2007iot,Jeet2015roa,Yamaguchi2015esf,Borisyuk2015bsa}.

    % alternative: 233U alpha decays, irrespective of the optical details (wavelength, lifetime, cross section, polarizations, flux, beamtime...)
The first four of these five challenges can be circumvented by an alternative approach to populate the isomeric state, namely through alpha decay $^{233}\mathrm{U} \rightarrow {}^{229\rm{m}}\mathrm{Th}$. Following a cascade of gamma transitions, about 2\% of the $^{229}$Th nuclei end up in the isomeric state \cite{Browne2008nds}. In the absence of competing decay channels, these nuclei will eventually de-excite into the nuclear ground state under emission of the sought-after VUV gamma. The wavelength of this gamma can be measured with a spectrometer. Such a spectroscopy experiment could reduce the present uncertainty in the transition wavelength to a degree that would allow one to commence laser spectroscopy.

    % experiments with 233U and MgF2 recoils so far
A number of experiments using $^{233}$U recoils have already been performed \cite{Richardson1998upe,Irwin1997ooe,Browne2001sfd,Moore2004sfa,Inamura2009sfa,Kikunaga2009hle,Burke2010ttb}, at least two are currently ongoing \cite{Zhao2012oot,Peik2013coo,Wense2013tad,Wense2015dot}. These experiments are designed such that the detection is well-separated in both space and time from the population of the isomer, as the latter process is accompanied by radioactivity and the associated radioluminescence. Commonly, a thin sample of $^{233}$U is brought into the vicinity of a UV-transparent crystal (e.g.~MgF$_2$ or CaF$_2$ ), such that $^{229}$Th recoil nuclei may leave the surface of the $^{233}$U source plate and deposit onto, or penetrate slightly into, the absorber plate.

    % In this paper, we...
In this Letter, we investigate an alternative approach, briefly mentioned already in Ref.~\cite{Hehlen2013oso}: $^{233}$U-doped crystals as a source of nuclear VUV gamma emission. The gamma flux of such crystals may be many hundred times larger compared to surface-implanted $^{229}$Th recoils, but the radioactive decay $^{233}\mathrm{U} \rightarrow {}^{229\rm{m}}\mathrm{Th}$ and the isomer gamma emission $^{229\rm{m}}\mathrm{Th} \rightarrow {}^{229\rm{g}}\mathrm{Th}$ are separated in neither space nor time. This strategy seems futile at first sight, as an enormous radioluminescence background is introduced into the detection volume.  We will show, however, that radioluminescence caused by $^{233}$U alpha decay does not overlap with the anticipated wavelength range around 160\,nm. Working towards the interpretation of optical spectra, we find that contaminations of the crystal can generate spectrally narrow features in the UV range, easily misinterpreted as the nuclear isomer signal.

	% how current experiments work
\textit{A bright source of $^{229\mathrm{m}}$Th} --- A number of current experiments use thin foils of $^{233}$U, often in the form of UO$_2$, as a source of $^{229}$Th nuclei. An energy of  $Q_{\alpha} = 4.91$\,MeV is released upon the $\alpha$-decay of $^{233}$U \cite{nudat}, where the $^{229}$Th nucleus obtains a recoil energy of up to 84\,keV. A Th ion of this kinetic energy has a penetration depth of about 15\,nm in UO$_2$, so if the event occurs close to the surface, and the direction of propagation leads towards the surface, then the Th ion might leave the substrate. The Th ion is caught on a large-bandgap absorber plate. After a certain time of accumulation, the absorber crystal is moved into a detector. This approach thus allows to separate the $^{233}$U alpha decay and the detection of the VUV gamma in both space and time.
%, where MgF$_2$ is usually chosen for its large bandgap

	% intrinsically limited flux
This method, however, has two severe limitations: At first, the flux of $^{229}$Th nuclei that can be implanted into the absorber plate is intrinsically limited by the small range of $^{229}$Th recoils in the UO$_2$ material, and the comparatively long half-life of the $^{233}$U isotope. The maximum flux of isomeric $^{229{\mathrm {m}}}$Th recoil nuclei per unit surface is
\begin{equation}
\Phi = C \, \ell \times \frac{\rho_{{\rm UO}_2}}{{\rm M_{UO_2}}}  \times N_A \times \frac{{\rm ln(2)}}{t_{1/2}} \times R \times B,
\end{equation}
where $\rho_{{\rm UO}_2}$ is the density of the UO$_2$ and M$_{{\rm UO}_2}$ its molar mass, $t_{1/2} = 159\,200\,$a the half-life of $^{233}$U, $\ell \approx 15\,$nm the range of 84-keV recoil ions, and $N_A$ is Avogardro's constant. The geometrical factor $C = 1/4$ acounts for the fraction of nuclei up to a depth of $\ell$ that reach the absorber plate, and $R$ is the probability of radiative de-excitation. For $R=1$ and a branching ratio of $B = 2\,\%$, the maximum gamma emission rate is $\Phi_{\gamma} \approx 26 / ({\rm s} \times {\rm cm}^{2})$.

	% uncertainty: recoils may not reach the bulk
The second experimental limitation stems from the fact that the penetration of the $^{229}$Th ions into the absorber plate is also only a few 10\,nm. Depending on the surface roughness and cleanliness, the recoil ions might be stopped before reaching the true crystal bulk structure. The band gap of the surface region of the crystal might be smaller than the energy of the isomeric state, allowing it to de-excite via electronic states \cite{Tkalya2000dot,Karpeshin2007iot}.

	% the new approach
To overcome these two limitations, we here follow a different approach:~ Doping the $^{233}$U directly into a suitable crystal will allow to increase the production rate of Th recoil ions, and the Th ions will be born directly into the bulk of the lattice. The flux of isomer gammas that can be extracted from the crystal is intrinsically limited by the optical absorption of uranium defect centers in the crystal. As we will show later, the optical absorption length is equal to $\xi_{\rm U}(\lambda)/n_{\rm U}$, where $n_{\rm U}$ is the uranium doping concentration (in terms of uranium nuclei per crystal unit cell) and $\xi_{\rm U}(\lambda)$ is a wavelength-dependent material constant. The isomer gamma flux reads
\begin{equation}
\Phi_{\gamma} = \xi_{\rm U}(\lambda) \times \frac{\rho_{\rm crystal}}{{\rm M_{crystal}}}  \times N_A \times \frac{{\rm ln(2)}}{t_{1/2}} \times R \times B,
\end{equation}
where ${\rm M_{crystal}}$ denotes the molar mass of a crystal unit cell. Assuming $R=1$  and the crystal thickness $d$ to exceed the absorption length, $d \gg \xi_{\mathrm U}(\lambda)/n_{\rm U}$, we obtain $\Phi_{\gamma} \approx 4200 / ({\rm s} \times {\rm cm}^{2})$ for $^{233}$U:CaF$_2$ \cite{SM}. Note that this approach allows for continuous signal integration and does not require knowledge of the isomer lifetime.

	% linking paragraph
To validate our approach, we will now quantify the optical transmission and measure the radioluminecence spectrum of our crystal of choice, which is U:CaF$_2$.

\begin{figure}
\includegraphics[width=\columnwidth]{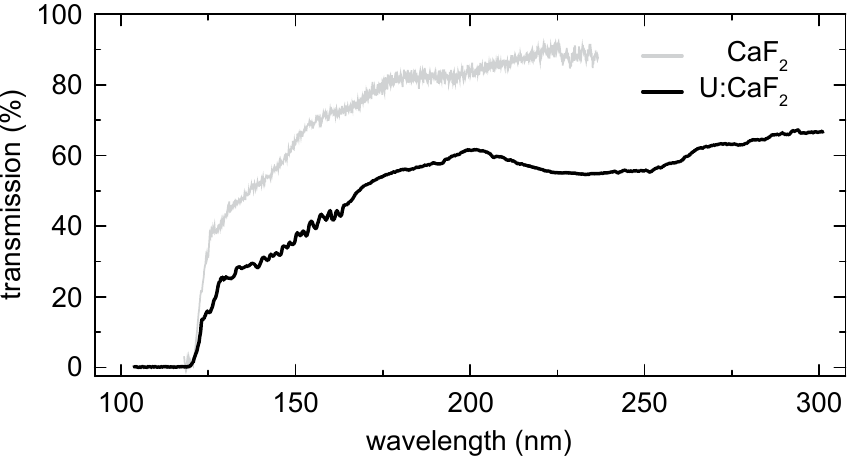}
\caption{Transmission of $^{238}$U:CaF$_2$. The black curve was taken with a crystal of 4.1\,mm thickness and $5 \times 10^{-5}$ doping concentration, the grey curve shows the transmission of an undoped specimen.}
\label{fig:fig1}
\end{figure}

	%Introduction, crystal preparation
\textit{Transparency of U:CaF$_2$ in the UV} --- Although U:CaF$_2$ is widely used as a laser material, transmission curves in the UV spectral range have not been available so far \cite{Hargreaves1992oso,Su2004iom}. We use an in-house furnace to produce a set of U:CaF$_2$ crystals with doping concentrations between $n_{\rm{U}} = 4.9 \times 10^{-5}$ and $2.0 \times 10^{-3}$, corresponding to uranium densities of $1.2 \times 10^{18}\,\rm{cm}^{-3}$ to $4.9 \times 10^{19}\,\rm{cm}^{-3}$, where we assume a crystal unit cell to be formed by one Ca and two F ions. Due to the radioactivity of $^{233}$U and its limited availability, we use depleted $^{238}$U as a chemically identical proxy of $^{233}$U for the studies presented here. The crystals have a ruby red color, indicating that the uranium atoms are in the trivalent state \cite{Su2004iom}. The crystals are cut and polished into discs of a few mm thickness and 17\,mm diameter.

	%Transmission measurement		the crystal used here: 4.1mm thickness, from V042_9, lowest doping: 2.5 mg UF4 in the entire crystal.
Transmission measurements are performed with a VUV spectrometer held at a pressure of $2\times10^{-6}$\,mbar, where a deuterium lamp is used for calibration and to generate the probe light. Figure \ref{fig:fig1} shows the transmission curve of a crystal with thickness $d=4.1\,\rm{mm}$ and doping concentration $n_{\rm{U}}=5 \times 10^{-5}$. 
	
	% absorption scales linearly with doping concentration
We model the transmission as
\begin{equation}
I(\lambda) = I_0 \,(1-a)\,e^{-  d (1/ \xi_{\rm {crystal}}(\lambda) + n_{\rm U} / \xi_{\rm {U}}(\lambda))},
\end{equation}
where $a$ describes losses at the crystal surfaces, $\xi_{\rm {crystal}}(\lambda)$ accounts for the absorption of an undoped crystal, and $\xi_{\rm {U}}(\lambda)$ describes the additional absorption due to the uranium doping. We find that the wavelength-dependent absorption coefficient $\xi_{\rm {U}}(\lambda)$ is independent of the doping concentration for values of $5 \times 10^{-5} < n_{\rm{U}} < 2 \times 10^{-3}$:~the absorption length scales inversely proportional with $n_{\rm{U}}$. For a wavelength of 160\,nm, we measure $\xi_{\rm U}(160\,{\rm nm}) = 6.3(8)\times10^{-5}\,{\rm cm}$. It is encouraging to find that this value is 40 times larger than its equivalent quantity $\ell$ in the approach of recoil implantation.

\begin{figure}
\includegraphics[width=\columnwidth]{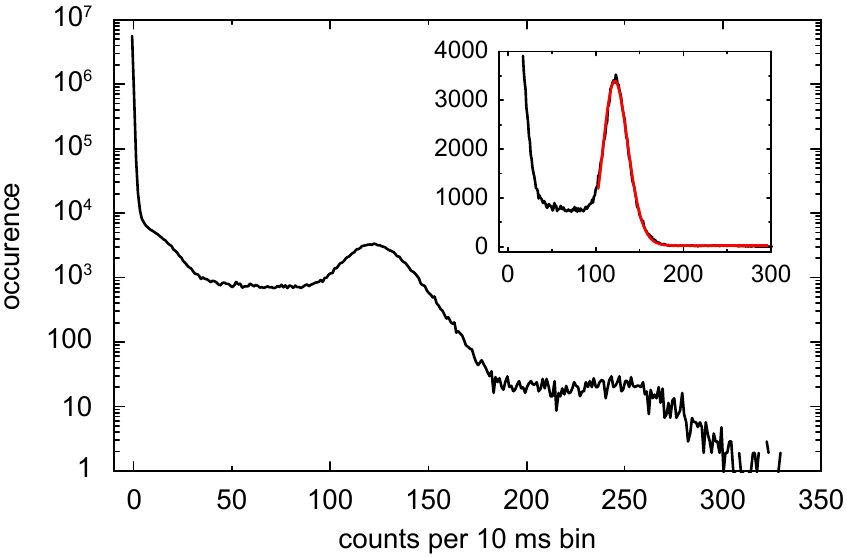}
\caption{Histogram of radioluminescence emission from a $^{238}$U-doped crystal. The main feature around 120 counts corresponds to a flash of UV photons following $\alpha$-decay, shown also in the inset with a Gaussian fit to the data (red line).}
\label{fig:fig2}
\end{figure}

	%burst measurement
\textit{Radioluminescence of U:CaF$_2$} --- A $^{238}$U:CaF$_2$ crystal with an alpha activity of 1.9\,Bq is placed in front of a Cs-Te PMT (sensitivity range 115 to 320\,nm). We record the emission of characteristic bursts of photons, where the rate of the bursts corresponds to the alpha activity of the $^{238}$U. Each burst lasts for a few \textmu s and contains some $10^5$ photons \cite{Stellmer2015rap}. We use a bin width of 10\,ms and plot the number of counts per bin in a histogram, shown in Fig.~\ref{fig:fig2}. A prominent feature around 120 counts is observed; this feature is associated with the alpha decay of $^{238}$U into $^{234}$Th. This isotope quickly decays into $^{234}$U (half-life 245\,000\,a) via two beta decays; these beta decays are reflected in the hump at about 10 counts.

	%temperature measurement
The radioluminescence emission rate $R$ is temperature-dependent \cite{Rodnyi1997ppi}. We measure a near-linear dependence $R(T) = (1-c\,(T-T_0))\, R(T_0)$ between 10 and $80\,^{\circ}$C, and obtain a slope of $c = 0.0107(5)\,\mathrm{K}^{-1}$ for $T_0 = 10\,^{\circ}$C. Mild heating from room temperature to $80\,^{\circ}$C thus reduces the radioluminescence already by a factor of almost four. From this characteristic temperature-dependence \cite{Rodnyi1997ppi}, as well as the signature histograms discussed above, we conclude that the radioluminescence properties of U:CaF$_2$ are governed entirely by the general properties of CaF$_2$ and are not determined by the uranium doping.

The half-life of $^{238}$U is too long, and the light throughput of optical spectrometers too small, to obtain a radioluminescence spectrum from $^{238}$U:CaF$_2$ crystals. As $^{233}$U:CaF$_2$ crystals are not yet available, we simulate such a crystal by bringing a thin layer of $^{233}$U (age 45(5) years, $^{232}$U contamination 12(1) ppm) in physical contact with a commercial CaF$_2$ sample (diameter 25\,mm, thickness 5\,mm). The $^{233}$U had been deposited onto a steel backing using electro-deposition \cite{Kazakov2014pfm}, the layer has a diameter of 22\,mm, a thickness of about 15\,\textmu m, and an activity of 7.5\,MBq (1500-times larger compared to the $^{229}$Th:CaF$_2$ crystals used in a related study \cite{Stellmer2015rap}). Note that only the top 15\,nm of the uranium layer give a flux of Th recoil ions into the CaF$_2$ sample, but the entire thickness contributes $\alpha$- and $\gamma$-particles.

\begin{figure}
\includegraphics[width=\columnwidth]{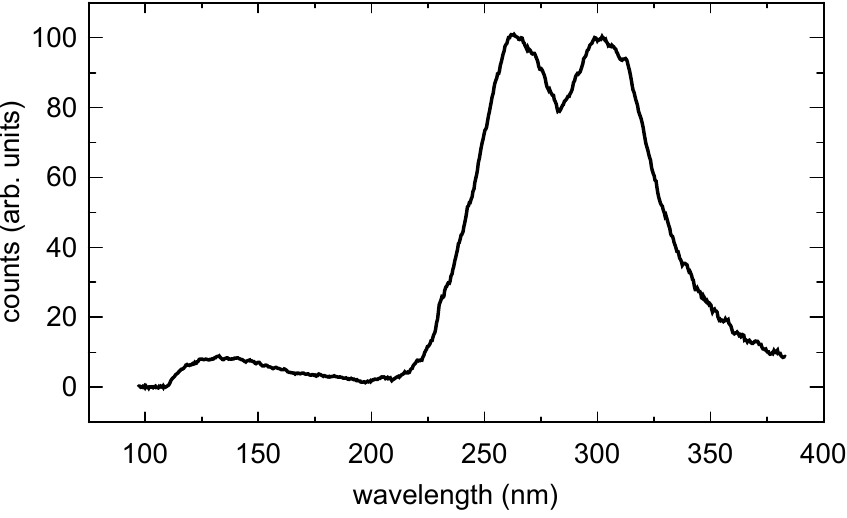}
\caption{Radioluminescence of CaF$_2$ in contact with $^{233}$U, showing scintillation between 220 and 400\,nm caused predominantly by $\alpha$-decay. The Cherenkov radiation, caused predominanly by $\beta$-decay of thorium daughters, extends down to the transmission edge of the crystal around 125\,nm.}
\label{fig:fig3}
\end{figure}

This stack is placed into the spectrometer. The luminescence spectrum, obtained after an integration time of 50 hours is shown in Fig.~\ref{fig:fig3}. The spectrum can be decomposed into two parts:~the scintillation of CaF$_2$ around 280\,nm \cite{Stellmer2015rap}, and Cherenkov radiation below 200\,nm. The spectral shape and amplitude of the scintillation feature depend on the specific type of crystal \cite{SM}. In CaF$_2$, it extends down to 220\,nm. This finding is very promising, as the scintillation is spectrally far away from the expected isomer wavelength at 160\,nm. The scintillation is caused primarily by alpha particles penetrating the crystal up to a depth of about 15\,\textmu m.\cite{SM}.

The broad spectral component between 120 and 200\,nm is attributed to Cherenkov light, caused by beta decay of $^{228}$Th and $^{229}$Th daughters. The cut-off at low wavelengths coincides the lower transmission edge of the CaF$_2$ crystal. The Cherenkov light is intrinsically broad and will constitute a locally flat background in the search for the spectrally very narrow isomer signal. Chemical purification of the $^{233}$U material prior to crystal fabrication can reduce the Cherenkov light substantially \cite{SM}. Note that the amplitude of Cherenkov radiation is proportional to the ingrowth of $^{228}$Th and $^{229}$Th; this dependence is exactly the same for doped and surface-implanted ions. The signal-to-background ratio of these two approaches is thus identical and depends only on the degree of $^{232}$U contamination and on the time elapsed since the last chemical removal of thorium ingrowth from the $^{233}$U source material. 

For a time of 100 days since chemical purification, and a $^{232}$U contamination of 10\,ppm, we calculate an emission of $1.9 \times 10^{-3}$ Cherenkov photons within a 1.0-nm wavelength window around 160\,nm for every $^{233}$U decay \cite{SM}. This number needs to be compared to the probability $R\times B$ of VUV gamma emission. Assuming $R=1\%$, $B=2\%$, $\Phi_{\gamma} = 4200 / ({\rm s} \times {\rm cm}^{2})$ and parameters of a standard VUV spectrometer, 1.5 days of measurement time are required for the isomer signal to exceed the noise of the background \cite{SM}.
	
\begin{figure}
\includegraphics[width=\columnwidth]{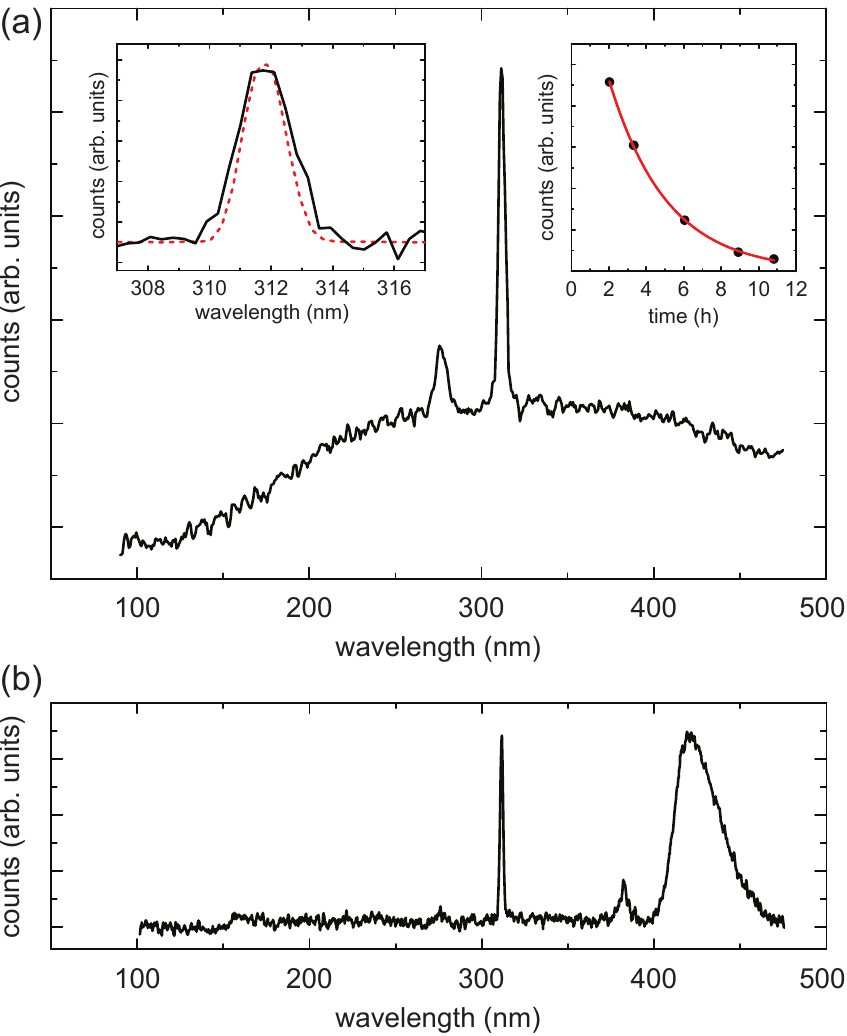}
\caption{Long-lived and narrow-bandwidt h scintillation of CaF$_2$ caused by contaminations. (a) Luminescence spectrum of CaF$_2$ upon irradiation with gamma radiation. The narrow feature around 312\,nm has a linewidth of less than 1.2\,nm and a lifetime of 3.1(1) hours; see the text for details. (b) The same narrow feature can be induced by irradiation with VUV light; the feature at 420\,nm is caused by F-centers.}
\label{fig:fig4}
\end{figure}

	% spectrally narrow features
\textit{Spectrally narrow features} --- While Cherenkov radiation is intrinsically free of narrow features, crystal contaminations could potentially cause spectrally narrow luminescence. To probe for such disturbing signals, we subject various CaF$_2$ samples to intense gamma radiation inside a shut-down nuclear reactor for 16 hours, and measure their luminescence spectra afterwards.

The emission of one specific commercial CaF$_2$ specimen is shown in Fig.~\ref{fig:fig4} (a). On top of a spectrally broad background, we observe two narrow features, the dominant one being located at 312\,nm. We monitor the amplitude of this narrow feature over time, and find an exponential decay with a time constant of $\tau = 3.1(1)$ hours (right inset of Fig.~\ref{fig:fig4} (a)). The decay of the broad background proceeds on a very similar timescale. The  spectral width of this feature is measured to be 2.05\,nm (left inset), largely limited by the instrumental resolution (red dashed line). Subtracting the signal width and the experimental resolution in quadratures, we obtain a natural linewidth of below 1.2\,nm.

The appearence of narrow and long-lived features in the luminescence spectrum of crystals is striking, as typical crystal defects show broad features of typically 10\,nm width, and a much faster, non-exponential decay. We attribute the observed feature to crystal contamination with Gd$^{3+}$ ions, which show narrow emission at a wavelength of 312\,nm \cite{Peik2015privatecomm}. We speculate that crystal defects induced by the gamma radiation slowly transfer their energy onto Gd$^{3+}$ ions, which radiate on a narrow transition. The same narrow feature can be provoked by illumination with a deuterium lamp \cite{Stellmer2015rap}; see Fig.~\ref{fig:fig4} (b).

Many current experiments search for a spectrally narrow feature with a lifetime on the order of an hour, and use both the linewidth and lifetime to discriminate the isomer signal from the crystal luminescence background. As shown above, contaminations of the crystal can easily be mistaken for the sought-after signal. This pitfall can be circumvented by performing the experiment with different types of crystals. Experiments building on the ${}^{233}\mathrm{U} \rightarrow {}^{229\mathrm{m}}\mathrm{Th}$ approach could perform reference measurements with  $^{234}$U, which is very similar to $^{233}$U in terms of half-life (245\,000\,a vs 160\,000\,a), $Q_{\alpha}$ (4.86\,MeV vs 4.91\,MeV), and half-life of the first daughter nuclei (75\,000\,a vs 7932\,a).

	% conclusion
To conclude, we proposed a new method to detect and measure the VUV isomer gamma of $^{229}$Th in consequence of $^{233}$U alpha decay in VUV-transparent crystals. We found that radioluminescence induced by the alpha decay is spectrally separated from the expected isomer wavelength region. This region, however, is covered by Cherenkov radiation, induced predominantly by the beta decay of $^{228}$Th and $^{229}$Th daughters. The flux of isomer gammas extracted from the crystal is so large that, even if the probability of radiative de-excitation of the isomer is only 1\%, the signal can be discriminated from the broad Cherenkov radiation after a spectroscopy measurement time of only a few days.

We thank J.~Schwestka, V.~Schauer, and V.~Rosecker for preparation of uranium samples. We greatly appreciate fruitful discussions with E.~Peik, L.~von der Wense, and P.\,G.~Thirolf. This work has received funding from the ERC project 258604-NAC and from the European Union's Horizon 2020 research and innovation programme under grant agreement No 664732 ``nuClock''.

\bibliographystyle{apsrev}

\bibliography{bib}

\setcounter{figure}{0}
\setcounter{table}{0}
\setcounter{equation}{0}
\renewcommand{\thefigure}{S\arabic{figure}}
\renewcommand{\thetable}{S\Roman{table}}
\renewcommand{\theequation}{S\arabic{equation}}

\newpage

\centerline {\large \textbf{SUPPLEMENTAL MATERIAL}}

\section{Cherenkov radiation}
\label{sec:Sec1}

\subsection{Overview}

The isomer transition is expected at a wavelength of 160(10)\,nm, where our approach of $^{233}$U-doped crystals is sensitive to the wavelength region between the UV transparency cutoff (125 nm for CaF$_2$) and the onset of the dominant scintillation features (220 nm in CaF$_2$, 380 nm in MgF$_2$). This spectral region is covered by Cherenkov radiation; see Fig.~3. The spectral shape of Cherenkov emission is essentially a convolution of the crystal's transmission, given by $\xi(\lambda)$, and its refractive index $n(\lambda)$, both of which change very smoothly with wavelength. Its amplitude is independent of the specific details of the crystal structure. Cherenkov radiation thus poses a locally homogeneous background rather than a sharp peak that could interfere with the sought-after isomeric signal.

Cherenkov radiation is caused by charged particles (in our case, electrons) travelling through the crystal at velocities $v$ larger than the speed of light in the medium, $v > c/n(\lambda)$, where $n(\lambda)$ is the wavelength-dependent index of refraction. In CaF$_2$, we find $n(160\,\mathrm{nm}) = 1.55$. The minimum threshold energy required for electrons to emit Cherenkov radiation is
\begin{equation}
E = m_e c^2 (\gamma - 1)
\end{equation}
with the electron rest mass $m_e$ and the relativistic parameter $\gamma = 1/\sqrt{1-\beta^2}$. Here, $\beta = v/c = 1/n(\lambda)$, and we obtain a minimum energy $E_{e\mathrm{, min}}$ of 158\,keV. Very similar values are obtained for other types of crystals.

The spectrum of the Cherenkov radiation is well described by the Frank-Tamm formula \cite{Jelley53cr},
\begin{equation}
\frac{dN_{ph}}{d\lambda \, dx} = \frac{2 \pi \alpha}{\lambda^2}\left(1-\frac{c^2}{v^2 n^2(\lambda)}\right), \label{eq:FrankTamm}
\end{equation}
where $dN_{ph}$ is the number of Cherenkov photons emitted in the wavelength range $d\lambda$ while the electron travels the elementary path $dx$ in the medium. $v$ is the speed of the electron, and $\alpha=e^2/(\hbar c)\simeq 1/137$. To obtain the total number of photons emitted, it is necessary to integrate over the path of the electron. Expressing the path $x$ in terms of the instant kinetic energy $E_e$, the integration reads
\begin{equation}
\frac{dN_{ph}}{d\lambda}(E_{e,0}) = \int\limits_{E_{e,min}}^{E_{e,0}}
\frac{dN_{ph}}{d\lambda \, dx} (E_e) \left| \frac{dx}{dE_e} \right| dE_e, \label{eq:FrankTammTotal}
\end{equation}
where $E_{e,0}$ is the initial energy of an electron. This calculation is straigforward using tabulated values of electron ranges $\ell_e(E_{e,0})$ in CaF$_2$ \cite{ESTAR} and the correspondence $x(E)=\ell_e(E_{e,0})-\ell_e(E)$.

\subsection{Origin of Cherenkov radiation and yield of various processes}

Electrons of large enough energy may originate from the following processes: (i) $\beta^-$~decay with sufficiently large energy $Q_{\beta}$ released, (ii) highly energetic conversion electrons (CE) accompanying radioactive transformation of nuclei, and (iii) highly energetic gammas, which interact with the crystal via the photoelectric effect (predominantly below 100 keV), Compton scattering (100\,keV to 10\,MeV) and pair production (above 10\,MeV). We will now look at these processes more closely and estimate the yield of Cherenkov radiation.

(i) {\em $\beta$-electrons} are characterized by a continuous energy spectrum, which may be described by the Fermi law \cite{Venkataramaiah85ff}
\begin{equation}
\begin{split}
\frac{dN_{\beta}}{dE_{e,0}}(E_{e,0}) = C\, F(z,E_{e,0})\sqrt{E_{e,0}^2+2E_{e,0} m_ec^2} \times \\
(Q_{\beta}-E_{e,0})^2 (E_{e,0}+m_ec^2),
\end{split}
\label{eq:Fermi}
\end{equation}
where $C$ is a normalization constant and $z$ is the atomic number of the daughter nucleus. The Fermi function $F(z,E_{e,0})$ describes the Coulomb interaction between the emitted electron and the nucleus; see Ref.~\cite{Venkataramaiah85ff} for an explicit expression. The average Cherenkov yield of $\beta$-electrons obtained in a decay with end-point energy $Q_{\beta}$ is given by
\begin{equation}
\frac{dN_{ph,\beta}}{d\lambda}(Q_{\beta}) = \int\limits_{E_{e,min}}^{Q_{\beta}}
\frac{dN_{ph}}{d\lambda}(E_{e,0})\frac{dN_{\beta}}{dE_{e,0}}(E_{e,0}) dE_{e,0},
\label{eq:BetaCher}
\end{equation}
where $\frac{dN_{ph}}{d\lambda}(E_{e,0})$ is given by Eq.~\ref{eq:FrankTammTotal}.

(ii) {\em Conversion electrons} are characterized by their discrete energy spectrum. Therefore, their yield may be calculated directly by formula \ref{eq:FrankTammTotal}.

(iii) {\em High-energy gammas} accompanying radioactive decay of nuclei may produce secondary high energetic electrons via Compton scattering, photoabsorbtion, and pair production. In the experiment considered here, the most energetic gammas are below 3\,MeV in energy. On the other hand, gammas with energy below $E_{e,min}=158$\,keV cannot produce electrons of sufficiently high energy to contribute to the Cherenkov radiation. The most efficient scattering process between a few 100\,keV and a few MeV, and the only one considered here, is Compton scattering.

The energy $E_{e,0}$ of a scattered electron is connected with the energy $E_{\gamma}$ of the incident gamma and the scattering angle $\theta$ as
\begin{equation}
1- \frac{E_{e,0}}{E_{\gamma}} = \frac{1}{1+\frac{E_{\gamma}}{m_e c^2} (1+\cos\theta)}. 
\label{eq:Compt}
\end{equation}

To find the minimum gamma energy $E_{\gamma}$ required to generate an electron with energy $E_{e\mathrm{, min}} = 158\,$keV (at the threshold to emit Cherenkov radiation), we set $\theta = 180^{\circ}$ and obtain $E_{\gamma\mathrm{, min}} = 295\,$keV.

The differential cross section of Compton scattering into the elementary energy of the scattered electron may be given with help of the Klein-Nishina differential cross-section into the elementary solid angle and relation \ref{eq:Compt} between the scattering angle and the energy of the scattered electron. The cross section reads
\begin{equation}
\begin{split}
\frac{d\sigma}{dE_{e,0}}&(E_\gamma, E_{e,0})=\, \pi r_e^2 \frac{m_e c^2}{E_\gamma^2}
\left[ 
\frac{E_\gamma'}{E_\gamma}+\frac{E\gamma}{E_\gamma'}+\right. \\
& \left. \left(\frac{m_e c^2}{E_\gamma'}-\frac{m_e c^2}{E_\gamma} \right)^2-2 \left(\frac{m_e c^2}{E_\gamma'}-\frac{m_e c^2}{E_\gamma} \right)
\right], 
\label{eq:ComptCrossSect}
\end{split}
\end{equation}
where $r_e=e^2/(m_ec^2)$ is the classical electron radius, and $E_\gamma'=E_\gamma-E_{e,0}$ is the energy of the scattered gamma.

The range of high-energy gammas in the crystal medium is much larger than the actual size of our crystals. For example, the probability of a $300$~keV gamma to interact with the CaF$_2$ crystal over a path of 5\,mm length is only about 15~\%; this value is even smaller for gammas of higher energy \cite{NIST}. This fact allows us to neglect the attenuation of gammas, as well as the interaction of scattered gammas with the medium. Assuming a path length of $\ell_{\gamma}=5$\,mm in the crystal, we estimate the average yield of a single gamma with energy $E_\gamma$ born in the crystal as
\begin{equation}
\begin{split}
&\frac{dN_{ph,Compt}}{d\lambda}(E_\gamma)=n_e \ell_\gamma \times \\
&\hphantom{aaa}
\int\limits_{E_{e,min}}^{E_{e,max}(E_\gamma)} \frac{d\sigma}{dE_{e,0}}(E_\gamma, E_{e,0}) 
\frac{dN_{ph}}{d\lambda}(E_{e,0})\, dE_{e,0}. 
\label{eq:ComptCher}
\end{split}
\end{equation}
where $E_{e,max}(E_\gamma)$ is given by Eq.~\ref{eq:Compt} at $\theta=180^\circ$.

\begin{figure}
\includegraphics[width=\columnwidth]{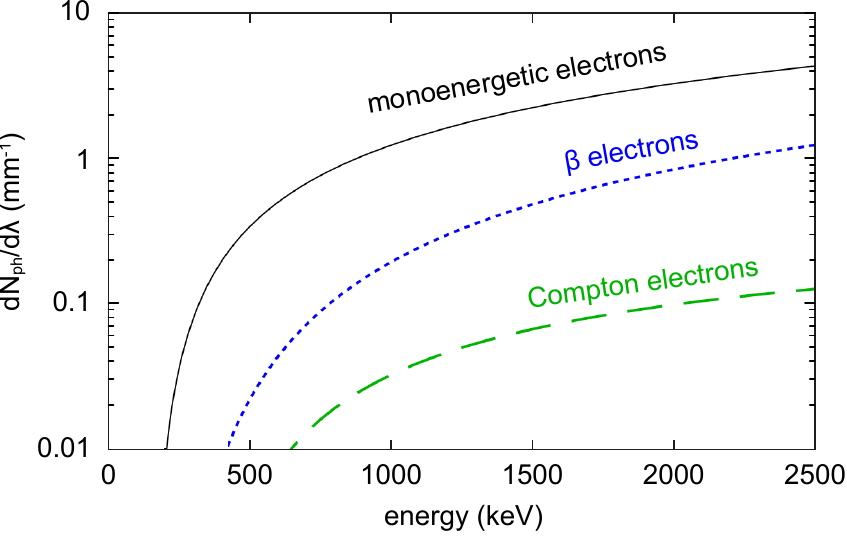}
\caption{Comparison of the average number of Cherenkov photons emitted within a 1-nm wide spectral window near $\lambda=160$\,nm on average by one monoenergetic electron of energy $E_{e,0}$ (black, solid), of $\beta$-electrons emitted in a decay with $Q_\beta$  (blue, dotted), and Compton electrons emerging from scattering with gammas of energy $E_\gamma$ (green, dashed). The path length of gammas in the crystal has been taken as $\ell_{\gamma}=5$\,mm.}
\label{fig:fig5}
\end{figure}

The Cherenkov yields of the three different processes discussed above are compared in Fig.~\ref{fig:fig5}.

\subsection{Ingrowth of daughters and contamination with $^{232}$U}

The Cherenkov radiation shown in Fig.~3 cannot be caused directly by the alpha decay of pure $^{233}$U, as this decay is not accompanied by a significant number of conversion electrons or gammas of sufficiently high energy. Instead, the Cherenkov radiation originates from activity in the sequence of short-lived $^{229}$Th daughters, namely (i) the beta decay of $^{225}$Ra, $^{213}$Bi, and $^{209}$Pb, (ii) conversion electrons, and (iii) the high-energy gammas of various daughters, e.g.~$^{213}$Bi.

The $^{233}$U source used for the measurements was not purified, and daughters have been building up for the past 45 years (measured by inductively coupled plasma mass spectrometry, ICP-MS). Chemical purification of the $^{233}$U is expected to be capable of removing 99.5\% of the thorium. Such a preparation would reduce the Cherenkov radiation initially by a factor of 200, with a very slow build-up over the $^{229}$Th lifetime.

The $^{233}$U material was obtained by neutron irradiation of $^{232}$Th through the steps $^{232}\mathrm{Th} + \mathrm{n} \rightarrow {}^{233}\mathrm{Th} \overset{\beta}{\rightarrow}  {}^{233}\mathrm{Pa} \overset{\beta}{\rightarrow} {}^{233}\mathrm{U}$. Depending on the details of the breeding process, the $^{233}$U contains non-negligible amounts of $^{232}$U as a by-product of the neutron irradiation. The short half-life of $^{232}$U (70.6\,a) and its first daughter $^{228}$Th (1.9\,a) lead to a high activity of the entire chain, which includes strong gamma emission. Unlike the $^{233}$U chain, which ``pauses'' at $^{229}$Th for nearly 8000 years, the $^{232}$U chain proceeds down to the stable $^{208}$Pb on timescales comparable to the duration of the experiment.

Using gamma spectroscopy, the content of $^{232}$U in the $^{233}$U plate source used here has been determined to 11.6(1.0)\,ppm. At this level, the gamma radiation of the $^{232}$U chain clearly dominates over the $^{233}$U chain, the same is true for beta decays. It is thus important that the spectroscopy experiment proposed in the main text be carried out with $^{233}$U material almost free of $^{232}$U.

\section{Realistic parameters for a spectroscopy experiment}

We will now attempt to model a future spectroscopy experiment. We assume a $^{233}$U:CaF$_2$ crystal of density $\rho_{{\rm CaF}_2} = 3.18\,$g/cm$^3$, thickness $d = 5\,$mm and  $^{233}$U doping concentration  $n_{\mathrm U} = 5 \times 10^{-4}$; such concentrations can easily be achieved. Note that here and in the main text, $n_{\mathrm U}$ denotes the amount of uranium ions per crystal unit cell, where, for simplicity, we assume a unit cell to contain one Ca ion and two F ions, $M_{{\rm CaF}_2} = 78.075\,$g/mol. The crystallographic unit cell would contain four Ca ions and eight F ions. Further, we assume the contamination of $^{232}$U to be $C_{232} = 10\,$ppm, and we assume that $t = 100\,$days have elapsed since the last removal of thorium (both $^{228}$Th and $^{229}$Th) from the source. The  $^{228}$Th and $^{229}$Th chains are in secular equilibrium. Such a crystal of $10 \times 5 \times 2\,$mm$^3$ would have a $^{233}$U activity of 170\,kBq (identical to the source used in Ref.~\cite{Zhao2012oot}) plus 4\,kBq from the $^{232}$U chain. The crystal absorption parameter is taken as $\xi_{\rm{U}}(160\, \mathrm{nm}) = 6.3 \times 10^{-5}\,$cm.

Concerning the experimental set-up, we consider a standard VUV spectrometer, where the crystal is imaged onto the entrance slit of the spectrometer using 1:1 imaging optics with a solid angle coverage of 0.14\% (e.g.~$f/2 = 200\,$mm, mirror diameter $D= 30\,$mm) and a mirror reflectivity in the UV of 80\%. The entrance slit has a height $h$ of 8\,mm and a width $w$ of 330\,\textmu m, and is imaged onto the detector with a concave grating. The grating is assumed to have a diffraction efficiency of 20\%, and the quantum efficiency of the detector is 30\%. The total detection efficiency as the product of solid angle, mirror reflectivity, and grating and detector efficiency, is $D=6.7 \times 10^{-5}$. A CCD camera allows to continuously measure the entire spectral region of interest. A grating with 1200 grooves/mm translates the 330\,\textmu m slit width to 1.0\,nm spectral width. The spectrum, captured by a standard CCD detector (8\,mm height, 30\,mm width), would cover 90\,nm, conveniently matching the window between the CaF$_2$ transparency cut-off at 130\,nm and the onset of alpha-radiation induced radioluminescence at 220\,nm.

\subsection{Signal amplitude}

The expression for the flux $\Phi_{\gamma}$ given in the main text (Eq.~2) relies on a number of simplifications. We will show now that these simplifications are justified for reasonable parameters. At first, the factor $\xi_{\mathrm U}$ stems from an integration over all sources of isomer gammas,
\begin{equation}
\int_0^{d} \! n_{\rm U} \, e^{-n_{\rm U}x/\xi_{\mathrm U}(\lambda)} \, {\rm d}x,
\end{equation}
where we assume $n_{\rm U}$ to be constant across the sample. For $d = \infty$, the integration yields exactly $\xi_{\mathrm U}(\lambda)$. For the parameters chosen above, the integral is $0.981 \,\xi_{\mathrm U}(\lambda)$, and more generally, the approximation is justified for $n_{\rm U}d/\xi_{\mathrm U}(\lambda) \gg 1$.

Secondly, the integral is only one-dimensional, assuming all VUV gammas to propagate perpendicular to the crystal surface. In correct terms, the expression of Eq.~2 in the main text would read 
\begin{equation}
\begin{split}
&\frac{d \Phi_{\gamma}}{d\Omega}(\vartheta = 0^\circ)= \\
&\hphantom{aaa}
\frac{1}{4\pi} \xi_{\rm U}(\lambda) \times \frac{\rho_{\rm crystal}}{{\rm M_{crystal}}}  \times N_A \times \frac{{\rm ln(2)}}{t_{1/2}} \times R \times B,
\end{split}
\end{equation}
which would give a maximum flux of $d\Phi_{\gamma}/d\Omega(\vartheta=0)=340/\mathrm{(s\times cm^2 \times srad)}$. This geometric simplification is certainly justified, as we consider light collection optics placed far away from the crystal, capturing only photons within a small solid angle. For the parameters given above, the largest deviation from normal incidence is $\vartheta = 4.3^{\circ}$.
%{$3.6^{\circ}$ for 25-mm mirror}, $\vartheta = 4.3^{\circ}$ for 30-mm mirror

As a third simplification, we assume that light absorption is dominated by the uranium content and not by the absorption of the CaF$_2$ crystal itself. In this way, the expression in Eq.~2 becomes independent of the doping concentration $n_{\rm U}$ and crystal thickness $d$.

By far the largest uncertainty in $\Phi_{\gamma}$ stems from the uncertainty in the probability $B$ to populate the isomer and the unknown magnitude of competing non-radiative decay channels. The possibility of non-radiative decay is captured in the quantity $R$, which denotes the probability of radiative decay. Using a values of $B = 2\%$ and $R=1$, we obtain $\Phi_{\gamma} = 4200 / (\rm{s} \times \rm{cm}^2)$. The rate of isomer gammas at the detector is 
\begin{equation}
\Phi_{\gamma}' = \Phi_{\gamma} \times h \times w \times D,
\label{Eq:DetectedRate}
\end{equation}
and we obtain $\Phi_{\gamma}' = 0.0075\,\rm{s}^{-1}$ for the experimetal specifications stated above.

In the following, we will compare this signal amplitude to various sources of background noise.

\subsection{Cherenkov radiation from $^{233}$U gammas}

\begin{table}
\begin{tabular*}{\columnwidth}{@{\extracolsep{\fill}}cccc}    
\hline\hline \noalign{\smallskip}
			 & & occurence    &  $\frac{dN_{ph}}{d\lambda}$ at 160~nm \\
$E_{e,0}$ [keV]     &       $\overline{E_{e,0}}$ [keV]      & per $10^6$      &  per $10^6$ \\    
                    &                             & disintegrations &  disintegrations\\
\noalign{\smallskip}\hline\noalign{\smallskip}
158-200             & 	178                       & 39.78           & 0.081     \\
200-300             & 	235                       & 68.25           & 1.8      \\
300-400             & 	314                       & 6.79            & 0.63     \\
$>$ 400               &   469                       & 0.0194          & 0.006   \\
\hline \hline
\end{tabular*}
\caption{Coarse-grained spectrum of conversion electrons with energies above $E_{e,min} = 158\,$keV that appear in the decay of $^{233}$U. Values are taken from Ref.~\cite{nudat}.}
\label{tab:233U_CE}
\end{table}

The alpha decay of $^{233}$U is accompanied by only very few conversion electrons and gammas of sufficiently high energy to generate Cherenkov light; see Tabs.~\ref{tab:233U_CE} and \ref{tab:233U_gamma_lines}. Using the models derived in Sec.~\ref{sec:Sec1}, we calculate that in total, there are on average only $N_{233} = 2.6\times 10^{-6}$ Cherenkov photons emitted into a 1-nm interval at $\lambda=160$~nm per disintegration of $^{233}$U. Comparing the emission of Cherenkov photons and isomer gammas, 
\begin{equation}
\frac{R_{\rm Ch,233}}{R_{\gamma}} = \frac{N_{233}}{R \times B},
\end{equation}
we find that for values $R > 10^{-4}$, the number of emitted isomer gammas exceeds the number of Cherenkov photons.

\begin{table}
\begin{tabular*}{\columnwidth}{@{\extracolsep{\fill}}cccc}    
\hline\hline \noalign{\smallskip}
 & & occurence    & $\frac{dN_{ph}}{d\lambda}$ at 160~nm \\
$E_{\gamma}$ [keV]   &   $\overline{E_\gamma}$ [keV]   & per $10^6$      &  per $10^6$ \\    
                    &                             & disintegrations &  disintegrations\\
\noalign{\smallskip}\hline\noalign{\smallskip}
295-350             & 	319                       & 118.6           &  0.0019    \\
350-500             & 	373                       & 10.4            & 0.0037    \\
500-750             & 	560                       & 0.6             & 0.033     \\
\hline \hline
\end{tabular*}
\caption{Coarse-grained spectrum of gammas with energies above $E_{\gamma\mathrm{, min}} = 295\,$keV that appear in the decay of $^{233}$U. Values are taken from Ref.~\cite{nudat}.}
\label{tab:233U_gamma_lines}
\end{table}

\subsection{Cherenkov radiation from beta decay in the $^{229}$Th chain}

\begin{table}[b]
		\begin{tabular*}{\columnwidth}{@{\extracolsep{\fill}}cccc}    
\hline\hline \noalign{\smallskip}
& & &  $\frac{dN_{ph}}{d\lambda}$ at 160~nm \\
decay   & $Q_{\beta}$ [keV]   &  occurence [\%]  & per 100 decays \\
        &                     &                  &  of $^{229}$Th \\
\noalign{\smallskip}\hline\noalign{\smallskip}
$^{209}\mathrm{Tl} \rightarrow {}^{209}\mathrm{Pb}$  & 1827        & 2.04  &  1.4   \\
$^{213}\mathrm{Bi}  \rightarrow {}^{213}\mathrm{Po}$  & 1423       & 64.8  &  27.5  \\
$^{213}\mathrm{Bi}  \rightarrow {}^{213}\mathrm{Po}$  & 983        & 30.2  &  5.5     \\
$^{209}\mathrm{Pb} \rightarrow {}^{209}\mathrm{Bi}$  & 644         & 100     & 5.4    \\
$^{225}\mathrm{Ra} \rightarrow {}^{225}\mathrm{Ac}$  & 356       & 31.2    & 0.13  \\
$^{225}\mathrm{Ra} \rightarrow {}^{225}\mathrm{Ac}$  & 316       & 68.8    & 0.14  \\
\hline \hline
		\end{tabular*}
    \caption{List of beta decays occuring in the decay chain of $^{229}$Th.}
    \label{tab:229Th_beta_decays}
\end{table}

There are four beta decays in the decay chain of $^{233}$U, one of which constitutes only a weak decay channel. From the ENSDF database \cite{ENSDF}, we extract the probability and released energy $Q_{\beta}$; see Tab.~\ref{tab:229Th_beta_decays}. All of the values of $Q_{\beta}$ are above $E_{e\mathrm{, min}} = 158\,$keV.

Next, we use the ESTAR values provided by NIST \cite{ESTAR} to calculate the path lengths of the electrons in CaF$_2$. Employing the continuous slowing-down approximation (CSDA), we obtain exemplary path lengths $s$ of $s(300\,\mathrm{keV}) = 0.33\,\mathrm{mm}$, $s(1\,\mathrm{MeV}) = 1.7\,\mathrm{mm}$, and $s(2\,\mathrm{MeV}) = 3.8\,\mathrm{mm}$. As a conservative simplification, we will assume that all electron paths are contained within the crystal.

We then employ Eq.~\ref{eq:BetaCher} to calculate the average number of Cherenkov photons within a 1-nm spectral window around 160\,nm, created along the entire chain of $^{229}$Th daughters. We find that 100 decays of $^{229}$Th are accompanied by the creation of 40 Cherenkov photons in the wavelength region of interest.

\subsection{Cherenkov radiation from gammas and conversion electrons in the $^{229}$Th chain}

As a next step, we go through the entire decay chain of $^{229}$Th down to $^{205}$Tl and, for each of the 10 decay steps, extract the gamma emission lines with energies above $E_{\gamma\mathrm{, min}} = 295\,$keV. Table \ref{tab:gamma_lines} lists all such gamma lines that have a probability of more than 0.1\% to appear in succession to a disintegration of $^{229}$Th. Only one gamma transition, located at 440.4\,keV, has a probability of more than a few percent. While we listed only transitions with probabilities above 0.1\% here, we checked that a summation over all the weaker transitions, as they appear e.g.~in the decay of $^{225}$Ac, can safely be neglected. 

An equivalent search is performed for conversion electrons (not listed here). We estimate the total contribution of conversion and Compton electrons to the Cherenkov radiation  background to be at the level of 0.74 and 0.15 photons per nm per 100 $^{229}$Th decays.\\

The Cherenkov photons are identical to the isomer gammas in wavelength, location of origin, and propagation to the detector. The ratio between the emission rates of Cherenkov photons vs isomer gammas for a decay of $^{233}$U reads 
\begin{equation}
\frac{R_{\rm Ch,229}}{R_{\gamma}} = \frac{t \times \lambda_{229}  \times N_{229} }{R \times B}.
\end{equation}
Here, $\lambda = \ln(2)/t_{1/2}$ is the decay constant. For $B=2\,\%$, $R=1$,  $N_{229} = 0.41$ Cherenkov photons in the wavelength region of interest, and a time of $t=100$ days allowed for the ingrowth of $^{229}$Th, we obtain a value of $4.9 \times 10^{-4}$.

\begin{table}
		\begin{tabular*}{\columnwidth}{@{\extracolsep{\fill}}ccc}    
\hline\hline \noalign{\smallskip}
decay   & $E_{\gamma}$ [keV]    & occurence [\%]\\
\noalign{\smallskip}\hline\noalign{\smallskip}
$^{209}\mathrm{Tl}  \rightarrow {}^{209}\mathrm{Pb}$  & 1566.9           & 2.08  \\
$^{213}\mathrm{Bi} \rightarrow {}^{213}\mathrm{Po}$  & 1100.2       & 0.27  \\
$^{213}\mathrm{Bi} \rightarrow {}^{213}\mathrm{Po}$  & 807.4        & 0.29  \\
$^{209}\mathrm{Tl}  \rightarrow {}^{209}\mathrm{Pb}$  & 465.1           & 2.02  \\
$^{225}\mathrm{Ac} \rightarrow {}^{221}\mathrm{Fr}$  & 452.2           & 0.11  \\
$^{213}\mathrm{Bi} \rightarrow {}^{213}\mathrm{Po}$  & 440.4        & 25.5  \\
$^{221}\mathrm{Fr} \rightarrow {}^{217}\mathrm{At}$  & 410.6         & 0.12  \\
$^{213}\mathrm{Bi} \rightarrow {}^{213}\mathrm{Po}$  & 323.7        & 0.16  \\
%$^{213}\mathrm{Bi} \rightarrow {}^{213}\mathrm{Po}$  & 292.8        & 0.42  \\
\hline \hline
		\end{tabular*}
    \caption{List of gamma lines with energies above  $E_{\gamma\mathrm{, min}} = 295\,$keV and a probability above 0.1\% to appear in the decay chain of $^{229}$Th.}
    \label{tab:gamma_lines}
\end{table}

\subsection{Cherenkov radiation from $^{232}$U contamination}

The production of $^{233}$U is plagued by the parasitic appearance of $^{232}$U. Even for contaminations at the ppm level, the activity of the material can be dominated by the $^{232}$U chain. The parent $^{232}$U undergoes alpha decay into $^{228}$Th (half-life 1.9\,a) without significant emission of high-energy gammas.

\begin{table}
		\begin{tabular*}{\columnwidth}{@{\extracolsep{\fill}}cccc}    
\hline\hline \noalign{\smallskip}
 & & & $\frac{dN_{ph}}{d\lambda}$ at 160~nm \\
decay   & $Q_{\beta}$ [keV]   &  occurence [\%]  & per 100 decays\\
        &                     &                  &  of $^{229}$Th \\
\noalign{\smallskip}\hline\noalign{\smallskip}
$^{212}\mathrm{Bi} \rightarrow {}^{212}\mathrm{Po}$  & 2252.1       & 55.3  & 57 \\
$^{208}\mathrm{Tl}  \rightarrow {}^{208}\mathrm{Pb}$  & 1801.3      & 17.7  & 12 \\
$^{212}\mathrm{Bi} \rightarrow {}^{212}\mathrm{Po}$  & 1524.8       & 4.50  & 2.19\\
$^{208}\mathrm{Tl}  \rightarrow {}^{208}\mathrm{Pb}$  & 1523.9      & 7.96  & 3.92 \\
$^{208}\mathrm{Tl}  \rightarrow {}^{208}\mathrm{Pb}$  & 1290.5      & 8.71  & 3\\
$^{208}\mathrm{Tl}  \rightarrow {}^{208}\mathrm{Pb}$  & 1038.0      & 1.14  & 0.23\\
$^{212}\mathrm{Bi} \rightarrow {}^{212}\mathrm{Po}$  & 739.4        & 1.44  & 0.12 \\
$^{212}\mathrm{Bi} \rightarrow {}^{212}\mathrm{Po}$  & 631.4        & 1.90 & 0.09 \\
$^{212}\mathrm{Pb} \rightarrow {}^{212}\mathrm{Bi}$  & 569.9        & 13.3  &0.42 \\
$^{212}\mathrm{Pb} \rightarrow {}^{212}\mathrm{Bi}$  & 331.3        & 81.7  &0.22 \\
\hline \hline
		\end{tabular*}
    \caption{List of beta decays occuring in the decay chain of $^{228}$Th with energies above 158\,keV and probabilities above 1\%.}
    \label{tab:228Th_beta_decays}
\end{table}

A list of beta decays in the $^{228}$Th chain is given in Tab.~\ref{tab:228Th_beta_decays}. We find that the decay of 100 $^{229}$Th nuclei is accompanied by 79 Cherenkov photons emitted by $\beta$-electrons in a 1\,nm interval near 160\,nm.

\begin{table}
		\begin{tabular*}{\columnwidth}{@{\extracolsep{\fill}}ccc}    
\hline\hline \noalign{\smallskip}
decay   & $E_{\gamma}$ [keV]    & occurence [\%]\\
\noalign{\smallskip}\hline\noalign{\smallskip}
$^{212}\mathrm{Bi} \rightarrow {}^{212}\mathrm{Po}$  & 1620.7       & 1.51  \\
$^{208}\mathrm{Tl} \rightarrow {}^{208}\mathrm{Pb}$  & 860.5        & 4.46 \\
$^{212}\mathrm{Bi} \rightarrow {}^{212}\mathrm{Po}$  & 785.4       & 1.11  \\
$^{212}\mathrm{Bi} \rightarrow {}^{212}\mathrm{Po}$  & 727.3       & 6.65  \\
$^{208}\mathrm{Tl} \rightarrow {}^{208}\mathrm{Pb}$  & 583.2        & 30.6 \\
$^{208}\mathrm{Tl} \rightarrow {}^{208}\mathrm{Pb}$  & 510.7        & 8.10 \\
$^{212}\mathrm{Pb}  \rightarrow {}^{212}\mathrm{Bi}$  & 300.1           & 3.18  \\
\hline \hline
		\end{tabular*}
    \caption{List of gamma lines with energies above  $E_{\gamma\mathrm{, min}} = 295\,$keV and a probability above 1\% to appear in the decay chain of $^{228}$Th.}
    \label{tab:228Th_gamma_lines}
\end{table}

We then take a look at the gammas emitted along the $^{228}$Th chain; see Tab.~\ref{tab:228Th_gamma_lines}. In analogy to the $^{229}$Th decay chain, we estimate the total contribution of conversion and Compton electrons to the Cherenkov background to be at the level of 0.75 and 5.4 photons per nm per 100 $^{228}$Th decays. In total, we have $N_{228} = 0.85$ Cherenkov photons emitted in a 1-nm interval near 160\,nm for every decay of $^{228}$Th, produced by all its daughters.

Assuming $t$ to be much shorter then the half-life of $^{228}$Th, the ratio of Cherenkov photons emitted per isomer gamma reads
\begin{equation}
\frac{R_{\rm Ch,228}}{R_{\gamma}} = \frac{t \times \lambda_{228} \times \frac{\lambda_{232}}{\lambda_{233}}  \times N_{228} \times C_{232}}{R \times B}
\end{equation}
and takes a value of 0.096 for $B=2\%$, $R=1$, an ingrowth time of $t=100$ days, and a contamination of $C_{232}=10$\,ppm.

\begin{table}
		\begin{tabular*}{\columnwidth}{@{\extracolsep{\fill}}ccccc}    
\hline\hline \noalign{\smallskip}
origin   & $\beta$ decay    & CE & Compton & $N$\\
\noalign{\smallskip}\hline\noalign{\smallskip}
$^{233}\mathrm{U}$ decay  & $-$       &  $2.5\times 10^{-6}$	&  $3.9\times 10^{-8}$ & $2.5\times 10^{-6}$\\
$^{229}\mathrm{Th}$ chain & 0.40        & $7.4\times 10^{-3}$	& $1.5\times 10^{-3}$ & 0.41\\
$^{228}\mathrm{Th}$ chain  &  0.79      &  $7.5\times 10^{-3}$ & 0.054 & 0.85\\
\hline \hline
		\end{tabular*}
    \caption{Summary of the amount of Cherenkov photons created in a 1-nm spectral window centered at 160\,nm, originating from beta decay, conversion electrons (CE), and Compton scattering. $N$ is the sum of these three processes.}
    \label{tab:Summary}
\end{table}

\subsection{Detector noise}

State-of-the-art CCD detectors can be cooled to $-100^{\circ}$C, which reduces the dark noise to a level that is entirely negligible in comparison to the read-out noise of typically less than one electron. Events related to the impact of highly energetic particles (e.g.~of cosmic origin or from environmental radioactivity) are the dominant disturbance. A routine to remove such events from the data set is required. 

We define a parameter $\Omega = t_{\rm exp} \times A$, where $t_{\rm exp}$ is the exposure time of a single image and $A$ is the binning area of the CCD chip. The binning area $A$ can be matched to the average area of ``cosmic'' events. We find that for $\Omega = 3.3\, \rm{s} \times \rm{cm}^2 $, about 5\% of all data points are contaminated by ``cosmic'' events; this value has proven to be a good choice for the detection of small signals. Using this value of $\Omega$, a sub-area of the CCD chip corresponding to a spectral width of 1.0\,nm could thus be integrated over 125 seconds. The isomer gamma signal integrated during this time (assuming $B=2\,\%$ and $R=1$) is comparable to the read-out noise.

\subsection{Measurement time and non-radiative decay probability}

Summing over all contributions, the ratio between Cherenkov photons emitted in the 1-nm window at 160\,nm and the isomer gammas reads
\begin{equation}
\frac{R_{\rm Ch}}{R_{\gamma}} = \frac{N_{233} + t \times (\lambda_{229} \,  N_{229} + \lambda_{228} \, \frac{\lambda_{232}}{\lambda_{233}} \, N_{228}\,  C_{232})}{R \times B}.
\label{Eq:SumOfCherenkov}
\end{equation}
Only for entirely unrealistic times $t$ smaller than a few minutes does the decay of $^{233}$U dominate over the two thorium chains, and only for unrealistically low concentrations $C_{232} < 0.5$ ppm does the $^{229}$Th chain dominate over the $^{228}$Th chain. We have thus identified the beta decays in the $^{232}$U contamination as the main source of Cherenkov radiation; refer to Tab.~\ref{tab:Summary} for a summary.

We will now calculate the integration time $T$ it takes for the integrated isomer signal to become comparable to the noise of the Cherenkov background,
\begin{equation}
\Phi_{\gamma}' \times T \simeq \sqrt{\Phi_{\gamma}' \times \frac{R_{Ch}}{R_{\gamma}} \times T},
\end{equation}
or
\begin{equation}
T \simeq \frac{1}{\Phi_{\gamma}'} \times \frac{R_{Ch}}{R_{\gamma}}.
\end{equation}
Taking the detected signal amplitude of $\Phi_{\gamma}'$ from Eq.~2 and Eq.~\ref{Eq:DetectedRate}, $R_{Ch}/R_{\gamma}$ from Eq.~\ref{Eq:SumOfCherenkov}, and assuming $t=100$ days, $C_{232}=10$ ppm, $B=2\%$, and a radiative decay probability of only $R=1\%$, we obtain a characteristic integration timescale of $T= 1.5$ days.

The comparatively short measurement time is very promising:~the isomer emission can be detected within a few days of measurement time even if only 1\% of the isomers undergo radiative de-excitation, or, equivalently, if $B$ was one order of magnitude smaller than expected. 

To give a second example, even for $R=10^{-3}$, a reduced contamination of $C_{232} = 3$ ppm, and a shorter ingrowth time of $t=30$ days, the isomer emission could be measured within a few weeks.

Note that, apart from the light throughput of the spectrometer, $t$ and $C_{232}$ are the only tunable parameters that have an effect on $T$. Note also that $T$ is independent of the spectrometer slit width, as the values of $N$ scale approximately linear with the spectral window defined by the slit width. Importantly, $T$ scales as $(R \times B)^{-2}$.

\section{Origin of the radioluminescence feature around 280\,nm}

The major fraction of $^{233}$U-induced radioluminescence in CaF$_2$ is radiated between 220 and 360\,nm, as shown in Fig.~3. We will now attempt to identify the kind of radiation that causes this scintillation. There are a number of candidates:~Th recoil ions implanted into the crystal, alpha and beta particles, gamma rays, as well as electrons and X-rays from secondary processes. To quantify the individual contributions, we place a commercial CaF$_2$ crystal (5\,mm thickness, 25\,mm diameter) in close proximity to a $^{233}$U sample. We insert three different absorbers in between the radiation source and the crystal and record the emission spectrum.

\begin{figure}
\includegraphics[width=\columnwidth]{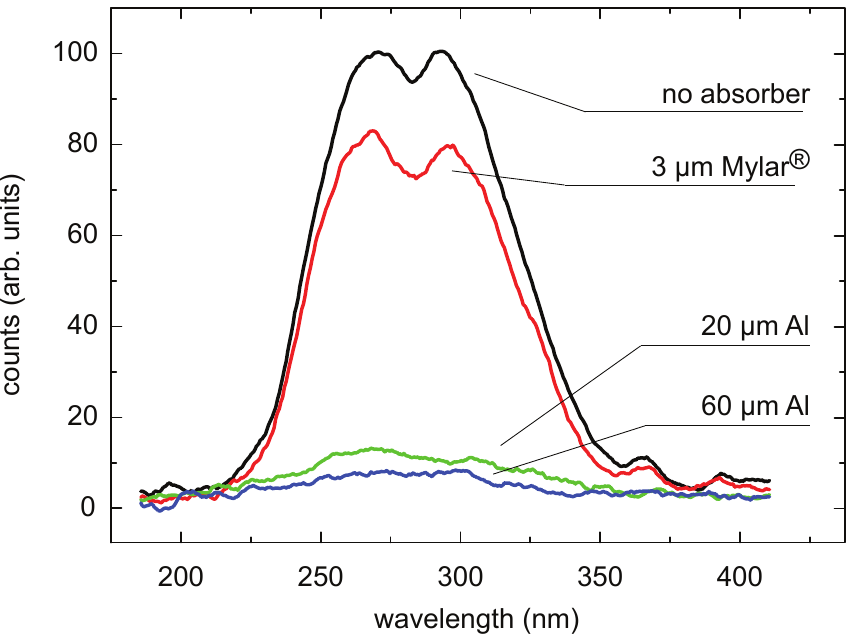}
\caption{Radioluminescence spectrum of CaF$_2$, brought into close proximity of a $^{233}$U source. Shielding the radiation by a thin Mylar$^{\circledR}$ foil (stops only Th nuclei) reduces the signal by 20\%, and shielding with aluminum foils (stop also alpha particles, yet transmits beta electrons and gamma rays) reduces the signal to below 10\%.}
\label{fig:fig6}
\end{figure}

	% Mylar foil
A 3-\textmu m Mylar$^{\circledR}$ foil securely absorbs the Th recoil ions, but transmits nearly all of the 4.8-MeV alpha particles (27\,\textmu m range) and all of the beta electrons, gamma rays, and X-rays. With this foil in place, we observe a 20\% drop in signal amplitude; see Fig.~\ref{fig:fig6}.

	% Al foil
We then substitute the Mylar$^{\circledR}$ foil by 20\,\textmu m of aluminum, which corresponds to the range of alpha particles. The foil transmits 99.4\% of all gamma rays at an energy of 30\,keV \cite{NIST} and all of the beta particles in question. The signal drops to 10\%, and it drops further to 5.5\% as the Al layer thickness is increased to 60\,\textmu m. 

	% Interpretation/Conclusion
The measurement suggests that the signal is caused predominantly by alpha particles, with much smaller contributions from recoil nuclei and gammas. This finding is consistent with the fact that more than 99\% of the energy deposited into the crystal from radioactive decay of our $^{233}$U source is via alpha particles. Note that the presence of the $^{232}$U chain increases the overall activity, but the partition of the energy released into alpha, beta, and gamma particles is roughly equal to the $^{233}$U chain.

 	% Exposure to high-energy gammas
To further study the effect of high-energy gamma radiation, we employ a more powerful source:~A commercial CaF$_2$ sample is placed directly into the core of a shut-down TRIGA Mark II reactor, where it is subjected to a massive flux of gamma rays. Alpha and beta particles are shielded by the water surrounding the fuel rods. After 15 hours of exposure, the crystal is quickly transferred into the spectrometer, but no signal resembling the spectrum of Fig.~\ref{fig:fig6} is observed. We conclude that, in absence of crystal contaminations, gamma radiation does not induce noticable long-lived defects.

	% Implantation of $\alpha$-particles
To complete our studies, we subject the CaF$_2$ sample to the radiation of a pure alpha emitter. We use a $^{241}$Am source with an activity of 5.5\,MBq (kindly provided by M.~Fugger, Atominstitut, Vienna), the sample is exposed to this radiation for 45 hours. The penetration depth of the 5.5-MeV alpha particles is estimated to be 15\,\textmu m. After an integrated bombardement of $2 \times 10^{11}$ alpha particles per cm$^2$, no damage or coloring of the crystal is observed, nor is the optical transmission reduced. A measurement of the luminescence spectrum begins 10 minutes after the end of exposure, but no signal of long-lived defects is observed between 120 and 500\,nm. A similar experiment with MgF$_2$ returns the same null result.

\section{Evaluating MgF$_2$ as a host crystal candidate}

\begin{figure}[h]
\includegraphics[width=\columnwidth]{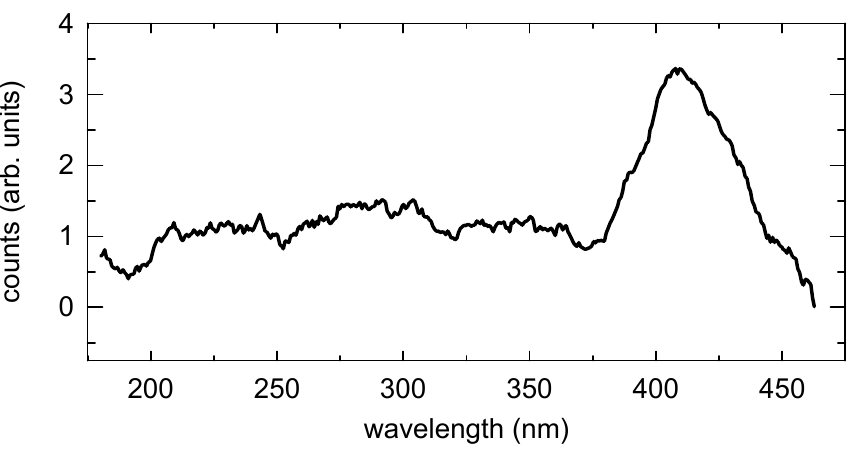}
\caption{Radioluminescence of MgF$_2$ in contact with $^{233}$U. A peak around 410\,nm is observed, with an amplitude much smaller compared to the spectrum of CaF$_2$, shown in Fig.~\ref{fig:fig5}. The units of the y-axes of these two graphs are identical.}
\label{fig:fig7}
\end{figure}

The radioluminescence of the $^{233}$U-doped crystal poses a considerable background for the detection of the isomer photon. The amplitude and spectrum of the radioluminescence emission depend on the specific crystal, and it might be worthwhile to consider several candidates. 

The Cherenkov spectrum is expected to be very similar for different types of crystal, as it depends only on the index of refraction (at 160\,nm, 1.55 for CaF$_2$, 1.47 for MgF$_2$). Only the scintillation induced by massive particles (alpha particles and recoiling ions) is expected to depend heavily on the specific crystal structure.

To assess the response of MgF$_2$ to the alpha decay, we attach the $^{233}$U layer to a MgF$_2$ specimen, just as we had done before with a  CaF$_2$ crystal. The radioluminescence spectrum is shown in Fig.~\ref{fig:fig7}, where the amplitude can be compared directly to the CaF$_2$ spectrum of Fig.~\ref{fig:fig6}. The pronounced feature around 300\,nm is missing, but instead, we observe a peak around 410\,nm. The amplitude of this peak is only about 3\% of the dominant peak in CaF$_2$. This drastic reduction in luminescence, appearing only at higher wavelengths, might be tempting for experiments with insufficient spectral resolution, e.g.~using photomultiplier tubes (PMTs). 

Crystals made up of light atoms (e.g.~LiF, MgF$_2$) are the preferred choice when X-rays are involved, as the X-ray absorption increases drastically with mass number. As an example, the mass attenuation coefficient of MgF$_2$ is 3.6-times smaller compared to CaF$_2$ for X-rays of 29\,keV energy \cite{NIST}. 

While uranium-doping of LiF has been reported \cite{Srinivasan1985pse}, we are not aware of any attempts to grow U:MgF$_2$ crystals. Measuring the doping efficiency of uranium into MgF$_2$ is the next step in the assessment of such crystals.

\end{document}